\documentclass[amsfonts, amssymb, amsmath, reprint, showkeys, nofootinbib, aps, pra, superscriptaddress]{revtex4-2}
\usepackage[latin9]{inputenc}
\usepackage{verbatim}
\usepackage{amsmath}
\usepackage{amssymb}

\usepackage{graphicx}

\makeatletter

\providecommand{\tabularnewline}{\\}

\usepackage{color,soul}

\makeatother

\begin{document}
\title{Realistic vulnerabilities of decoy-state quantum key distribution}
\author{Ivan Sushchev}
\email{sushchev.is16@physics.msu.ru}
\affiliation{SFB Laboratory, LLC, 127273 Moscow, Russia}
\affiliation{Quantum Technology Centre, Faculty of Physics, Lomonosov Moscow State
University, 119991 Moscow, Russia}
\author{Kirill Bugai}
\affiliation{SFB Laboratory, LLC, 127273 Moscow, Russia}
\affiliation{Bauman Moscow State Technical University, 2nd Baumanskaya Str. 5-1, Moscow, 107005, Russia}
\author{Sergei Molotkov}
\affiliation{Institute of Solid State Physics, Russian Academy of Sciences, Chernogolovka, Moscow Region 142432, Russia}
\author{Daniil Bulavkin }
\affiliation{SFB Laboratory, LLC, 127273 Moscow, Russia}
\author{Anna Sidelnikova}
\affiliation{SFB Laboratory, LLC, 127273 Moscow, Russia}
\author{Dmitry Melkonian}
\affiliation{SFB Laboratory, LLC, 127273 Moscow, Russia}
\author{Veronika Vakhrusheva}
\affiliation{SFB Laboratory, LLC, 127273 Moscow, Russia}
\affiliation{Quantum Technology Centre, Faculty of Physics, Lomonosov Moscow State
University, 119991 Moscow, Russia}
\author{Roman Lokhmatov}
\affiliation{SFB Laboratory, LLC, 127273 Moscow, Russia}
\author{Dmitriy Dvoretskiy}
\affiliation{SFB Laboratory, LLC, 127273 Moscow, Russia}
\affiliation{Bauman Moscow State Technical University, 2nd Baumanskaya Str. 5-1, Moscow, 107005, Russia}

\begin{abstract}
\section*{Abstract}
We analyze realistic vulnerabilities of decoy-state quantum key distribution
(QKD) arising from the combination of laser damage attack (LDA) and
unambiguous state discrimination (USD). While decoy-state QKD is designed
to protect against photon-number-splitting and beam-splitting attacks
by accurately estimating the single-photon fraction, it relies on
stable attenuation to prepare pulses with fixed mean-photon numbers.
An eavesdropper (Eve) can exploit LDA to irreversibly alter the optical
components on Alice\textquoteright s side, effectively increasing
the mean-photon numbers beyond the decoy-state security regime. We
show that once the alteration exceeds a critical threshold---on the
order of 10--20 dB---Eve can implement an efficient USD-based intercept-resend
strategy using current off-the-shelf technology, thus obtaining the
entire secret key. Numerical simulations confirm that for sufficiently
elevated mean-photon numbers, Eve\textquoteright s conclusive measurement
outcomes skew the decoy-state statistics, yet remain undetected by
standard security checks. We further demonstrate how a modified USD
setup employing an additional beam splitter can reduce the required
threshold, facilitating Eve\textquoteright s attack.  Additionally, we introduce the pseudo-photon-number resolution (PPNR) USD attack, which allows Eve to emulate all observable gains at Bob\textquoteright s side so that she remains fully undetectable even with advanced statistical checks. Our findings
emphasize the need for robust safeguards against high-power laser
damage in QKD systems, including careful hardware selection, rigorous
testing under high-power illumination, and real-time monitoring to
ensure the integrity of the decoy-state protocol.
\end{abstract}
\maketitle

\section*{Introduction}

Quantum key distribution (QKD) is believed to provide a secure communication
guaranteed by the laws of quantum physics. For the majority of QKD
protocols the security proof is present for single-photon implementation.
However, real-world QKD systems usually operate with faint laser pulses,
since appropriate single-photon sources are not widely available.
This inconsistency is often resolved by the GLLP approach \cite{GLLP}
combined with the decoy-state method \cite{key-decoy}, which allow
legitimate parties (Alice and Bob) to produce a secret key, even though
the unwanted multiphoton component is present. QKD then becomes resilient
to multiphoton-component attacks, such as photon number splitting
(PNS) \cite{PNS}, beam splitting (BS) \cite{BS}, conditional beam splitting (CBS) \cite{CBS} and unambiguous
state discrimination (USD) \cite{USD,dusek2000unambiguous}. This, however, is only the
case when the proper operation of the decoy-state method is implied,
i.e. the mean-photon numbers in Alice's pulses are fixed (apart from the small fluctuations which can be incorporated into the security proof \cite{sixto2022security, kamin2025r}). In turn,
an eavesdropper (Eve) can launch the laser damage attack (LDA) \cite{LDA_bug1,LDA_bug2,LDA_mak1,LDA_mak2}
to alter the attenuation at Alice's side. This results in the increase in Alice's pulse intensities, which is not addressed by the decoy-state method. Thus, when the system is subject to LDA, the multiphoton attacks should be reexamined.


LDA is a well-known strategy to affect the mean photon numbers in
Alice's pulses by altering the attenuation of the fixed or variable
attenuators using high-power radiation \cite{LDA_bug1,LDA_bug2,LDA_mak1}.
It has been reported that the continuous-wave radiation with 0.3 -- 5~W mean power can reduce the attenuation coefficient by up to 10 dB and more for the fixed attenuators  \cite{LDA_bug1,LDA_bug2,LDA_mak1} and by up to 14.5 dB for the variable attenuators \cite{LDA_mak1}. It has been postulated that such alteration compromises the security
of the decoy-state QKD \cite{LSA}, making it vulnerable to PNS attack.
However, to complete PNS, Eve needs to handle quantum non-demolition
(QND) measurement of the photon number, save the quantum state into
the quantum memory and replace the quantum channel with a lossless
line. This is argued to be beyond the present-day technologies \cite{gis,mol},
although some progress has arisen in this area \cite{LDA_exp,LDA_raman}. This raises the question, what are the real risks assosiated with LDA?

Ideally, every imaginable security drawback of a QKD system should be assessed and eliminated. In reality, infinite security is unattainable and practical security analysis of real-world QKD systems should be performed~\cite{gis} (for example, such security analysis was recently presented in context of the Trojan-horse attack \cite{sushchev2024trojan}). LDA has been addressed in several research works posing themselves as a guide to certification procedure, such as ETSI White Paper \cite{ETSI}, and practical security analysis of a commercial QKD system \cite{qrate}. Among other things, ETSI White Paper postulates that LDA does not appear to be intrinsically more threatening than other attacks. In \cite{qrate} authors introduce the risk evaluation system of the attacks under investigation. Each loophole is evaluated by three criteria: (1)~loophole likelihood, (2)~future or current technology, and (3)~amount of key leakage. LDA got mark M (2 points out of 3) --- medium risk, because it requires research and possibly future technology to exploit. We, however, believe that the real risks associated with LDA are underestimated in these works. Therefore, by providing a realistic strategy for Eve to steal the whole key using present-day technologies in our paper we try to illustrate that LDA is among the most dangerous attacks, and the corresponding security risks are the highest.

Let us consider all possibilities for Eve to succeed in LDA and evaluate them as in \cite{qrate}. The straightforward way for her is to increase the radiation intensity to classical level (large number of photons), so that she can perform classical measurements and effortlessly extract the whole key. The reported values for attenuation alteration of a single attenuator do not exceed 14.5 dB, therefore, such opportunity appears unlikely and obtains 0+1+1=2 points. The ultimate strategy is to perform PNS or the full version of CBS using QND measurements and/or quantum memory. This obtains 1+0+1=2 points due to practical unattainability. Alternatively, Eve can use BS or more practical version of CBS without long-term qubit storage, but she will never obtain the whole key this way, as discussed in \cite{CBS}, so the result is 1+1+0=2. We also note that even this lighter version of CBS requires a large number of near-perfect single-photon detectors, as well as a lossless communication line, which seems hard to implement in practice.

Finally, Eve can perform the USD attacks. They allow Eve to steal the entire key using only linear optical elements and single-photon detectors \cite{USD, dusek2000unambiguous}. She also does not need to use a lossless channel. Such attacks have recenly been shown to compomise the security of the coherent-one-way QKD protocol \cite{rey2024hacking}. Bennet et. al \cite{BS} suggested a simple scheme to permorm USD by dividing the pulses into two halves and conducting measurements in different basis for each half (which is essentially a copy of Bob's setup). When three detectors click, Eve can be sure about the state sent by Alice. While this scheme is not optimal (see \cite{van2002unambiguous} for its comparison with other USD schemes), it is completely technologically available.
In this paper, we show that this USD strategy will work with realistic attenuation alterations, starting from some critical value, and give the whole key to Eve. Thus, all three criteria are not mutually exclusive for LDA and the total score is 1+1+1=3, which corresponds to the highests risks for decoy-state QKD systems.

\section*{Results}
\subsection*{USD attacks}
The decoy-state method is commonly employed to estimate the single-photon gain $Q_1$, i.e., the rate of detection events on Bob's side that originate from the single-photon component of Alice's emitted pulses. When operating properly, the decoy-state method guarantees security against multiphoton-component attacks, including USD, since the key is shortened during the privacy-amplification stage in accordance with the $Q_1$ estimation (Eqs.~(\ref{S})--(\ref{E1})).

Let us consider the attack against weak+vacuum decoy-state BB84 QKD (see Methods). First, Eve launches LDA and modifies mean photon numbers for the signal and decoy states as follows:
\begin{eqnarray}
\tilde{\mu} & = & \kappa\mu\label{eq:mu_LDA}\\
\tilde{\nu} & = & \kappa\nu\label{eq:nu_LDA},
\end{eqnarray}
where $\kappa$ is the attenuation alteration coefficient, determined by the particular high-power radiation effect on the attenuator under attack.

\begin{figure}[htbp]
\centering \includegraphics[width=1\linewidth]{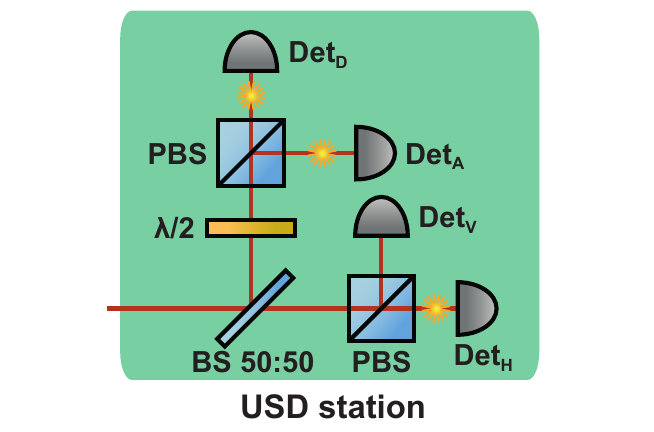} \caption{Eve's polarization-encoding setup for the USD attack. The depicted distribution of photons corresponds to a conclusive outcome for horizontally polarized state as an example. Det$_\text{H,V,D,A}$, detectors for horizontal, vertical, diagonal and anti-diagonal polarization; PBS, polarization beam-splitter; BS 50:50, symmetric beam-splitter; $\lambda/2$, half-wave plate for diagonal basis selection. The same setup can be designed for phase encoding.}
\label{fig:fig1} 
\end{figure}

Next, we consider the USD strategy without QND measurements and quantum memory, hence, achievable by today's technologies. In
fact, Eve only needs a copy of Bob's setup with passive basis choice for the full key distinction
(Fig. 1). Having a conclusive outcome, when strictly 3 detectors have clicked (meaning that the two clicks in one basis are wrong, and the single click in another basis is correct), she sends a corresponding classical pulse to Bob, so he detects it with near-100\% probability. The gains are then determined by the probability of Eve's conclusive measurements
outcome. Alternatively, Eve can employ a "translucent" USD variant, preserving part of Alice's pulse and forwarding it to Bob only when an unambiguous state-elimination measurement succeeds; in this case, two Eve's detectors in different bases must click so that the bit becomes known after basis reconciliation. This approach is likely comparable in effectiveness, but here we adopt the standard USD formulation to avoid dealing explicitly with channel transmittance and Bob's detection efficiency and later exploit its built-in state-selection capability in the PPNR USD attack.

It turns out that there exists the critical value for $\kappa$ where the decoy-state secret key
length estimation starts giving non-zero value, while the whole key leaks. This threshold $\kappa_{USD}$ can be found numerically. However, for small $\tilde{\mu}$ and $\tilde{\nu}$, the approximate analytical estimation takes place:

\begin{equation}
\kappa_{USD}=2+\frac{2}{\mu-\nu}\ln\frac{\mu}{\nu}=2\kappa_{3ph},\label{eq:an}
\end{equation}
where  $\kappa_{3ph}$ is the threshold for 3-photon PNS attack (see Methods). This relation underlines the connection between the USD attack and the straightforward, yet impractical, way to extract bits from 3-photon pulses, i.e., 3-photon PNS attack. The main difference between them is that the conclusive outcomes in USD sometimes arise from $n>3$ photon components (for example, when two photons arrive at the same detector). Consequently, Eve needs additional 3 dB to exceed this 3-photon threshold. This result illuminates that $n>3$ photon component gain complicates concealing the multiphoton attack.

\subsection*{Modified USD attack}

Eve can then modify her setup by adding a beamsplitter with transmittance $T$ to divert and detect "odd" photons --- those contributing to $n>3$ detection events, i.e., events requiring at least four photons (a conclusive USD outcome plus an extra click) --- and thus suppress yields from the $n>3$ components (Fig.~2). Although, the gains at Bob's side will drop, it will provide the reduced USD threshold value $\kappa_{T}$ (see the derivation in "Methods"):

\begin{equation}
\kappa_{T}=\frac{2}{2-T}\Big(1+\frac{1}{\mu-\nu}\ln\frac{\mu}{\nu}\Big)=\frac{2\kappa_{3ph}}{2-T}\label{kappa_T}
\end{equation}

Surprisingly, stronger blocking the multiphoton signals with $n>3$ even for conclusive outcomes increases the efficiency of USD attack and reduces the critical value~$\kappa_{T}$, meaning that the attack becomes succesfull at lower intensities. Asymptotically, $\kappa_{T}$ coincides with 3-photon attack limit $\kappa_{3ph}$ for $T\rightarrow0$. However, the gain $Q_{\mu}$ will drop significantly:

\begin{eqnarray}
\lim_{T\rightarrow0}Q_{\mu} = \frac{T^{3}}{32}e^{-\tilde{\mu}}\tilde{\mu}^{3}
\end{eqnarray}

Therefore, this strategy should be applied carefully, as gain drop
may cause a timeout error at Bob's side. 

\begin{figure}[htbp]
\centering \includegraphics[width=1\linewidth]{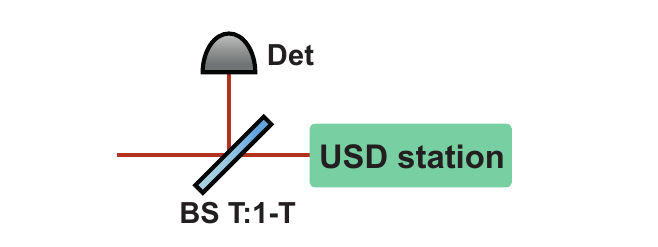} \caption{Eve's setup for the modified USD attack. Det, additional detector for ''odd'' photons monitoring; BS T:1-T, asymmetric beam-splitter. The fraction of the input radiation proportional to $T$ goes to the USD station, the remainder goes to Det; if Det fires, Eve forwards nothing to Bob.}
\label{fig:fig2} 
\end{figure}

\subsection*{USD attacks simulation results}

The yields for the described USD attacks are determined by the probabilities for Eve's conclusive measurements outcome for $n$--photon component. Considering such yields, we have simulated the decoy-state
single-photon gain estimation $Q_1$ over different attenuation alterations
$\kappa$ (Fig. 3). As seen, the estimated single-photon gain starts
to rise rapidly after exceeding the threshold. The critical $\kappa$
values correspond to those, derived by numerically solving Eq. \eqref{eq:tr}. In the worst studied case, corresponding to decoy-state parameters $\mu = 0.5$ and $\nu = 0.1$, the threshold lies near 11 dB, which can be further reduced below 10 dB by proper selection of the beam splitter transmittance $T$ in the modified USD. The reported experimental demonstrations of more than 10 dB attenuation alteration~\cite{LDA_mak1, LDA_bug2} thus provide strong evidence of the practical attainability of the proposed attack. Table I shows the comparison between these numerical
values with 0.1 dB rounding precision and analytical values from Eq. (\ref{eq:an}). There is a noticeable
difference around 1~dB between numerical and analytical values. However,
analytical estimations are conservative, which makes it acceptable
to use them during security analysis.

\begin{table}[b]
\caption{Critical $\kappa$ value (threshold) estimation by solving numerical
transcendental equation (\ref{eq:tr}) and from analytical equation
(\ref{eq:an}) for different decoy-state parameters }

\begin{tabular}{|c|c|c|}
\hline 
 & Numerical threshold & Equation (\ref{eq:an})\tabularnewline
\hline 
\hline 
$\mu=0.5$, $\nu=0.1$ & 11.1 dB & 10.0 dB\tabularnewline
\hline 
$\mu=0.5$, $\nu=0.01$ & 14.5 dB & 12.5 dB\tabularnewline
\hline 
$\mu=0.1$, $\nu=0.01$ & 18.3 dB & 17.2 dB\tabularnewline
\hline 
\end{tabular}
\end{table}

\begin{figure}[htbp]
\centering \includegraphics[width=1\linewidth]{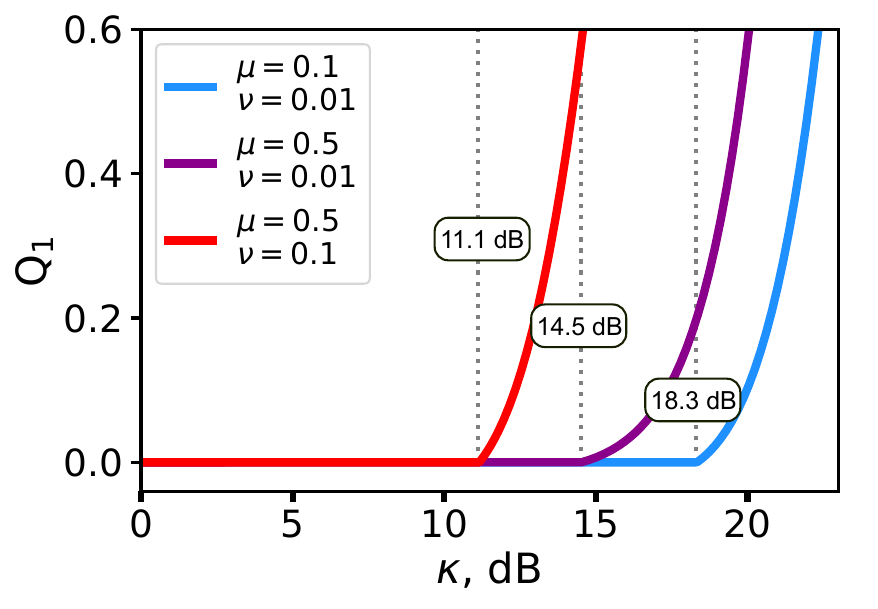} \caption{Dependence of the simulated single-photon gain estimation using Eq. (\ref{Q1}) on the attenuation alteration for different decoy-state parameters during the USD attack: 
 $\mu=0.5$, $\nu=0.1$ (red); $\mu=0.5$, $\nu=0.01$ (purple); $\mu=0.1$, $\nu=0.01$ (blue). The threshold values (11.1 dB, 14.5 dB and 18.3 dB, respectively) are indicated by the grey lines.}
\label{fig:fig3} 
\end{figure}

The rapid rise of the single-photon gain estimation in Fig. 3 gives
Eve the possibility to set the desirable $Q_{1}$ value by slightly
adjusting $\kappa$. The typical $Q_{1}$ values for proper functioning
QKD systems are dependent on the channel length, mean-photon number
$\mu$ and detection efficiency and lie in the region $10^{-6}-10^{-2}$.
The total gain $Q_{\mu}$ is also dependent on mentioned parameters
and usually should be of the same order of magnitude as $Q_{1}$.
Eve can reach this picking up the proper beam-splitter for the modified
USD. Operating with typical altered mean-photon numbers $\tilde{\mu}\approx5-15$
Eve can tune the gain to match the expected values at Bob's station by selecting the proper $T$
value (Fig.~4).

\begin{figure}[htbp]
\centering \includegraphics[width=1\linewidth]{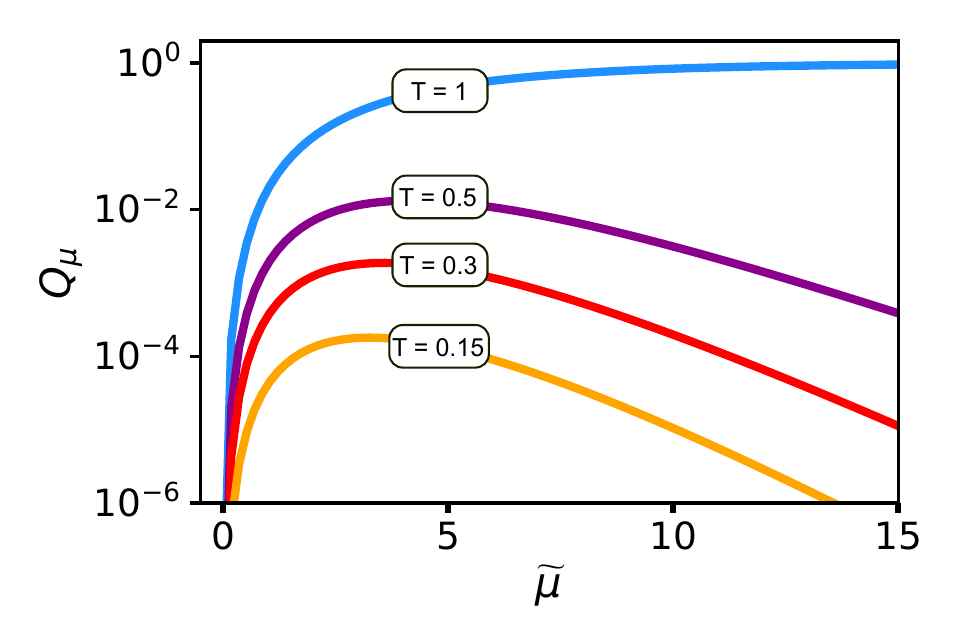} \caption{Simulated signal pulses gain dependence on the altered mean-photon number for modified USD attack with different beam-splitter transparency $T$: $T=1$ (blue); $T=0.5$ (purple); $T=0.3$ (red); $T=0.15$ (yellow).}
\label{fig:fig4} 
\end{figure}

\subsection*{Pseudo-PNR USD attack}

\begin{figure*}
\centering \includegraphics[width=0.8\textwidth]{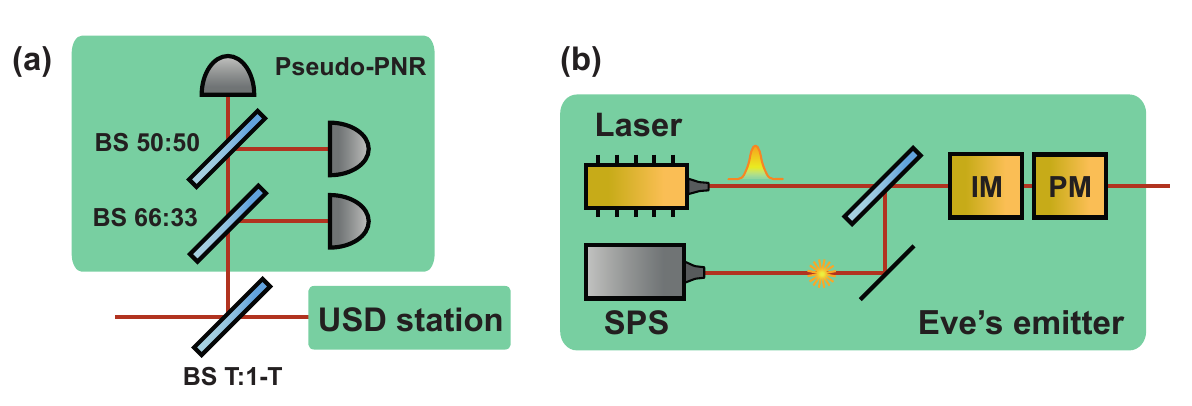} \caption{(a) Eve's PPNR USD measurement setup. (b) Eve's emitter: SPS, single photon source; IM, intensity modulator, PM, polarization modulator.}
\label{fig:fig5} 
\end{figure*}

In the general case, any intercept-resend attack that involves blocking (including PNS) alters the detection statistics observed by Alice and Bob. Niederberger et al. \cite{niederberger2005photon} showed that, in principle, Eve can reproduce the statistics of the original pulses while performing PNS attack. The decoy-state protocol, however, prevents this: blocking single photons produces non-proportional changes in the gains of signal and decoy pulses. Quantitatively, the single-photon gain $Q_1$---the target parameter of the decoy-state method---will therefore be affected. Above, we heve shown that the USD attack, and the modified USD attack combined with LDA, can nevertheless allow Eve to produce such gains that Alice and Bob incorrectly estimate $Q_1$ as a non-zero value. 

The decoy-state protocol targets two primary parameters: the single-photon gain $Q_1$ and the single-photon error probability $E_1$. However, a QKD system can, in principle, monitor additional statistical quantities and abort the session if any of them deviate from their expected values (for instance,the machine-learning methods can be employed to detect Eve using statistical anomalies \cite{xu2024automatically}). Moreover, numerical key-rate estimation methods cast the secret-key length calculation as an optimization problem and thus permit bounding Eve's information using the full set of measured parameters (see, e.g., Refs.~\cite{coles2016numerical, winick2018reliable}). Our USD and modified-USD strategies are designed solely to falsify $Q_1$ and are therefore, in principle, detectable by such additional checks. Below we construct a more sophisticated USD-based attack that employs pseudo-photon-number resolution (PPNR); this strategy enables Eve to reproduce every parameter observable at Bob's side, thereby evading detection based on the monitored statistics.

The following four parameters in addition to the error rates can be monitored by Bob in our weak+vacuum decoy-state BB84 protocol to detect Eve's presence:
\begin{itemize}
  \item $Q_\mu \approx \mu\eta\eta_B + p_{\mathrm{dcr}}$ -- gain for signal pulses;
  \item $Q_\nu \approx \nu\eta\eta_B + p_{\mathrm{dcr}}$ -- gain for decoy pulses;
  \item $Q_\mu^{(2)} \approx (\frac{\mu\eta\eta_B}{2})^2 + \mu\eta\eta_B\,p_{\mathrm{dcr}} + p_{\mathrm{dcr}}^2$ -- double-click gain in mismatch bases for signal pulses;
  \item $Q_\nu^{(2)} \approx (\frac{\nu\eta\eta_B}{2})^2 + \nu\eta\eta_B\,p_{\mathrm{dcr}} + p_{\mathrm{dcr}}^2$ -- double-click gain in mismatch bases for decoy pulses,
\end{itemize}
where $\,\eta\,$ is the channel transmittance (assumed known to Alice and Bob), $\eta_B$ is the quantum efficiency of Bob's detectors, and $p_{\mathrm{dcr}}$ is the dark-count rate.

Note that the gains for vacuum states do not change under any USD strategy, since Eve forwards a signal to Bob only when at least three of her detectors fire. If Eve's detectors have the same dark-count rate as Bob's, then the probability that Bob detects a pulse forwarded by Eve scales as $p_{\mathrm{dcr}}^3$, while Bob's own dark-count probability is $p_{\mathrm{dcr}} \gg p_{\mathrm{dcr}}^3$.

The setup for the PPNR USD attack is depicted in Fig.~5. The measurement setup is a variant of the modified-USD station augmented with a pseudo-PNR detector resolving up to three photons. Such a detector can be realized, for example, with one 50:50 and one 66:33 beamsplitter so that the probability for a photon to reach each output port is equal (see Fig.~1(a)). PPNR detectors can also be implemented in integrated optics architectures used for quantum computing~\cite{deng2023gaussian, skryabin2025heralded}. Each conclusive (3-click) outcome of the USD station is now accompanied by a PPNR measurement result that yields 0, 1, 2, or 3 photon counts. Eve associates these outcomes with four events --- single decoy-pulse click, double decoy-pulse click, single signal click, and double signal click, respectively --- and uses this association to guess which event is most likely to occur at Bob's station.

To impose her guess on Bob, Eve prepares the corresponding state using her emitter (Fig.~1(b)). Instead of sending a classical pulse, she now chooses between a laser pulse and a single photon. In the latter case, she must employ a single-photon source (SPS), which is less accessible than standard telecom lasers but can nevertheless be utilized by Eve~\cite{senellart2017high}. If the PPNR result corresponds to a single-click event (0- or 2-photon count) she sends a single photon from SPS to produce a click with controllable probabilities $p_{\nu,1}$~and~$p_{\mu,1}$ (adjusted via an intensity modulator, IM). If the PPNR result corresponds to a double-click event (1- or 3-photon count) she sends a weak coherent pulse with a controllable intensity $\nu'$~and~$\mu'$ (tuned by IM). In all cases Eve sets the polarization according to the USD outcome using a polarization modulator (PM).

\subsection*{Pseudo-PNR USD attack simulation results}

By selecting the proper values for her signals ($\nu'$, $\mu'$, $p_{\nu,1}$, and $p_{\mu,1}$) according to the equations (\ref{Q2_mu_eq})--(\ref{Q_nu_eq}), Eve can match all four parameters monitored by Bob with a small margin, as long as the certain threshold $\kappa_{PPNR}$ is exceeded. This threshold is depended on the communication channel losses and Eve's beamsplitter transparency $T$, as well as on  the quantum efficiency of Eve's and Bob's detectors, and the dark-count rate of Bob's detectors (the dark-count rate of Eve's detectors is negligible since Eve operates with relatively high intensities). 

\begin{figure}[htbp]
\centering \includegraphics[width=1\linewidth]{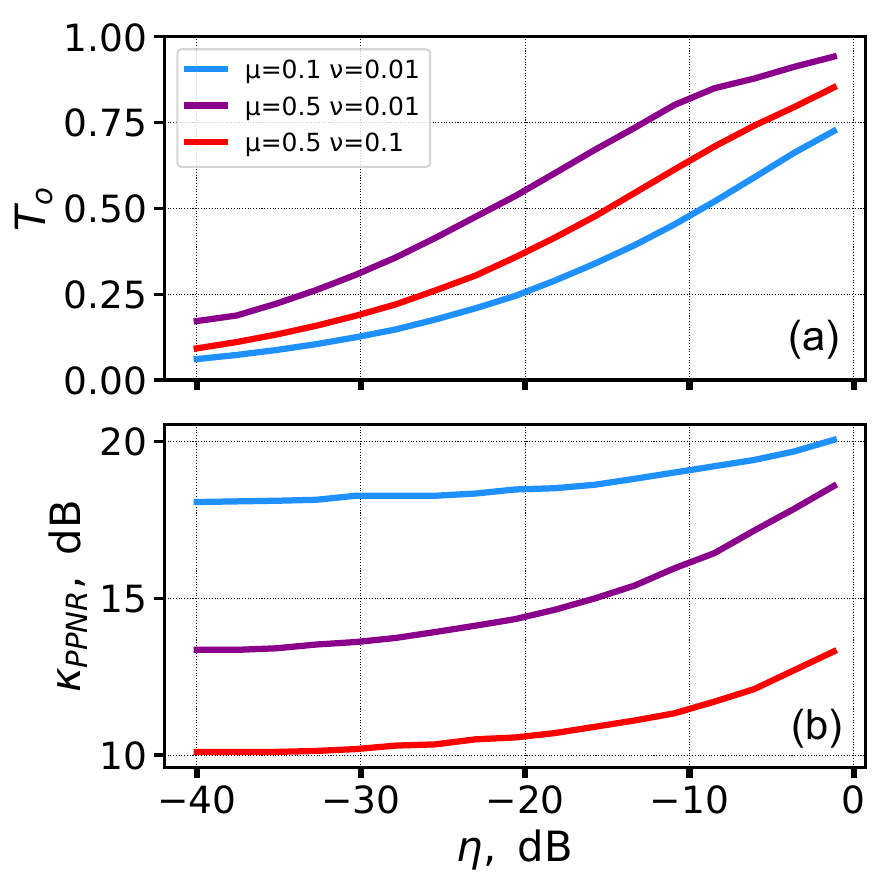} \caption{(a) The optimal value for the beamsplitter transmittance $T_o$ and (b) the threshold value for attenuation alteration for PPNR USD attack $\kappa_{PPNR}$ over the channel transmittance $\eta$ for different decoy-state parameters. The following parameter values have been used: $\eta_E=100\%$, $\eta_B=10\%$, $p_{dcr}=0$. The gains produced by the attack have matched the expected ones with a relative tolerance of $10\%$. The different colors in both figures represent different decoy-state parameters: 
 $\mu=0.5$, $\nu=0.1$ (red); $\mu=0.5$, $\nu=0.01$ (purple); $\mu=0.1$, $\nu=0.01$ (blue).}
\label{fig:fig6} 
\end{figure}

\begin{figure*}
\centering \includegraphics[width=1\textwidth]{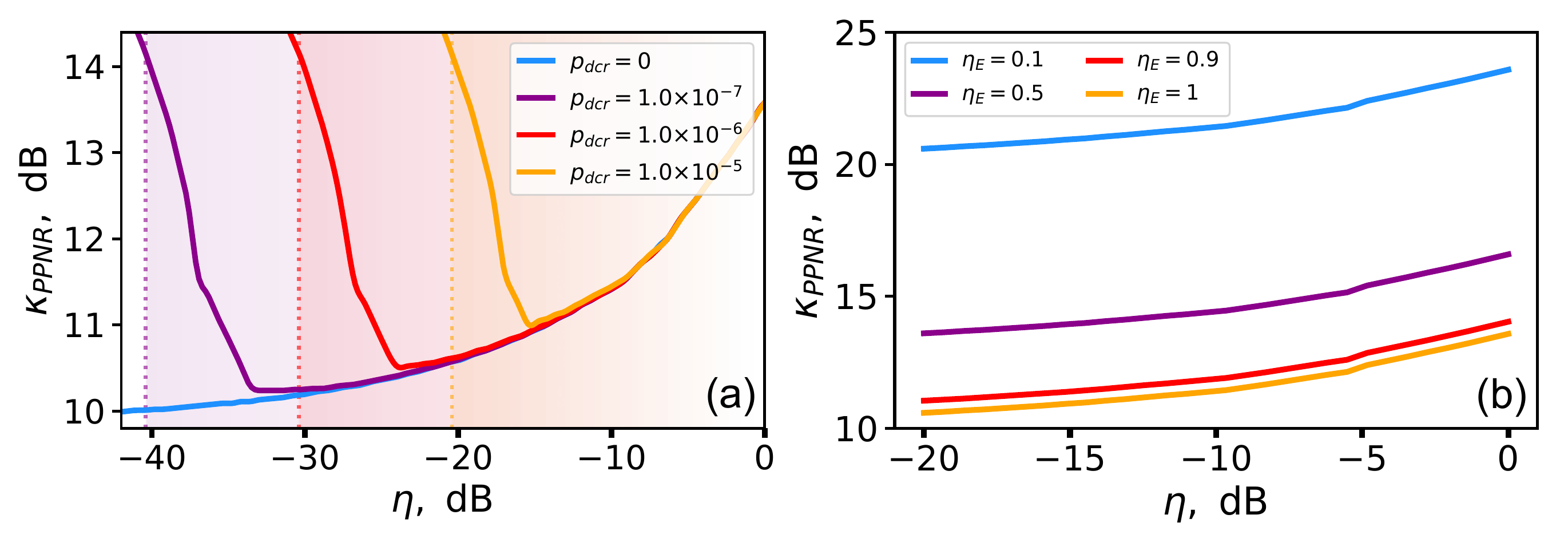} \caption{(a) The threshold attenuation alteration for PPNR USD attack $\kappa_{PPNR}$ over the channel transmittance $\eta$ for different dark-count rate of Bob's detectors ($p_{dcr}=0$ --- blue, $p_{dcr}=10^{-7}$ --- purple, $p_{dcr}=10^{-6}$ --- red, $p_{dcr}=10^{-5}$ --- orange); $\eta_E=1$, $\eta_B=0.1$; the dotted lines for each dark-count rate mark the critical losses for decoy-state BB84, the gradient colors  show the areas for each dark-count rate where positive key rates are possible. (b) The threshold attenuation alteration for PPNR USD attack $\kappa_{PPNR}$ over the channel transmittance $\eta$ for different quantum efficiency of Eve's detectors ($\eta_{E}=0.1$ --- blue, $\eta_{E}=0.5$ --- purple, $\eta_{E}=0.9$ --- red, $\eta_{E}=1$ --- orange). The decoy-state parameters are $\mu=0.5$, $\nu=0.1$.}
\label{fig:fig7} 
\end{figure*}

We have obtained the numerical dependence of the attenuation-alteration threshold $\kappa_{PPNR}$ on the channel transmittance $\eta$ for three sets of decoy-state parameters and for the beamsplitter transmittance $T_{o}$ that minimizes the corresponding threshold value. The results are shown in Fig.~6. One can see that, in the case of Bob's noiseless detection, additional losses make it easier for Eve to emulate the expected statistics. As seen in Fig.~6(b), the asymptotic $\kappa_{PPNR}$ values lie relatively close to the estimates obtained for our standard USD strategy (see Table~I). On the other hand, the $\kappa_{PPNR}$ values at zero loss are 2--3~dB higher than those values. A similar effect can be observed in Fig.~3 for the standard USD attack, where Eve also requires a few additional decibels of attenuation alteration for the estimated $Q_1$ to reproduce the expected statistics at low losses.

We have also run the simulations with the presence of the dark counts at Bob's detectors. In opposite to the ideal case, the threshold values grow starting from the certain amount of channel losses (see Fig.~7(a)). This is explained by the fact the dark counts statistics becomes significant at long distances and complicate Eve's attempts to reproduce the expected gain values. This, however, does not represent a serious obstacle for Eve, since the dark-count rate itself limits the maximum losses for QKD. According to~eq.~(\ref{E1}) the single-photon error rate is mainly determined by the signal-to-noise ratio for the decoy pulses. Conservatively considering the dark counts as the only source of errors, we can find the critical losses $\eta_{c}$ for which~11\% error-rate bound is reached:
\begin{equation}
\eta_{c} = \frac{p_{dcr}}{0.11\,\nu\,\eta_B}
\end{equation}
Therefore, the QKD session can take place only for $\eta~>~\eta_{c}$. These regions are highlighted in Fig.~7(a) for different dark-count rates. As seen, the threshold values $\kappa_{PPNR}$ begin to increase approximately 5~dB above the critical-loss point and rise by only a few additional decibels when approaching the upper bound.

Finally, we have studied the influence of Eve's detectors efficiency $\eta_E$ on the threshold value $\kappa_{PPNR}$. The simulated $\kappa_{PPNR}$ over channel transmittance $\eta$ for different $\eta_E$ is depicted Fig.~7(b). As seen, decreasing the efficiency results in proportional increase of the threshold. The reason of such behavior is that the imperfect detection effectively reduces the altered mean-photon numbers 
\begin{eqnarray}
\tilde{\mu} &\rightarrow \tilde{\mu} \cdot \eta_E = \mu\cdot \kappa\eta_E,\\
\tilde{\nu} &\rightarrow \tilde{\nu} \cdot \eta_E = \nu\cdot \kappa\eta_E.
\end{eqnarray}
When $\eta_E$ is small, Eve needs to increase $\kappa$ to reach the same altered mean-photon numbers and, consequently, the same threshold value. For example, for $\eta_E=0.1$ the threshold values rise by 10 dB.

\section*{Discussion}

We have described a realistic attack on QKD systems, which combines
the laser damage and the USD. We have presented Eve's USD setup, which
is essentially a copy of Bob's station. We have also presented the
setup for modified USD by adding an asymmetric beam-splitter, which
advances the efficiency of the attack. Unlike PNS or BS attack, such USD
can be implemented using present-day technologies. Finally, we have presented the PPNR USD attack, which allows Eve to reproduce all the parameters monitored by Bob, thus, completely evading detection.

Let us now discuss the practical limitations for Eve to implement the proposed attacks. First, we have assumed that Eve can place her laboratory near Alice's
station, therefore, we neglected the channel losses. If this is not the case and the signal must travel through, say, 10 km of standard optical fiber from Alice, the resulting channel loss would raise the threshold by approximately 2~dB. Second, one should take into account Eve's imperfect detection. We have determined that the efficiency of Eve's detectors $\eta_{E}$ raises the threshold value for the attenuation alteration (Fig.~7(b)). For typical InGaAs SPADs efficiency values $\eta_{E}=5-30\%$ it means that Eve needs additional 5--15 dB to reach the attack threshold. It then appears that the required attenuation alteration is significantly larger than the experimentally reported 10--14 dB. While for some QKD solutions this may seem like a safeguard, others often employ two attenuators in series (e.g., a variable and a fixed attenuator, as in Ref.\cite{qrate}), whose combined attenuation may fall even below this "pragmatic" threshold. Moreover, Eve can avoid these limitations for QKD systems operating
in the visible region (free-space QKD), where she can use high-efficiency Si SPADs.
She can also upgrade her apparatus by using superconducting detectors
with near-unity efficiency, which are available as a full-fledged device \cite{SNSPD}.

Another practical limitation for Eve is the deadtime of her detectors. For InGaAs SPADs, the deadtime is typically set to 0.1--10~$\mu$s to reduce the afterpulsing probability. This restricts the repetition rate of Eve's USD measurements to 100~kHz--10~MHz. If the QKD system operates at a higher repetition rate, Eve may wish to reduce the deadtime. This, however, would lead to parasitic afterpulsing detection events, which could introduce errors into Eve's measurements. To avoid this, Eve may employ state-of-the-art InGaAs SPADs with negligible afterpulsing probabilities and deadtimes as short as 3--10~ns~\cite{restelli2013single}. Alternatively, she could use superconducting detectors, which do not suffer from afterpulsing and have deadtimes of only a few nanoseconds~\cite{SNSPD}. Thus, considering the worst-case scenario, we can imply that Eve possesses sufficient resources to buy (or steal) state-of-the-art apparatus.



It is important to note that USD measurements allow Eve to identify the state without intrinsic errors. Nevertheless, Eve may still obtain an incorrect result or impose an erroneous state on Bob due to imperfections in state preparation and detection within the QKD system. As stated in Ref.~\cite{dusek2000unambiguous}, such effects do not require separate consideration, since USD attacks provide Eve with complete information about the key, even though Bob may register some errors. Moreover, it may even be advantageous for Eve to introduce a small additional QBER (for instance, by sending states orthogonal to her USD outcomes) to conceal her presence, as a nonzero QBER is typically observed even in the absence of any attack.

It is worth noting that, according to the reported experimental results \cite{LDA_mak1, LDA_bug2}, the attenuation alteration induced by LDA is inherently unpredictable. This leads to two possible scenarios for the adversary. In the first case, the alteration does not exceed the required threshold, and the attack fails. In the second case, the attenuation alteration exceeds the threshold, but its exact value remains unknown to Eve. The latter scenario is of primary interest in our analysis, as we focus on conditions under which the attack becomes feasible -- even if such conditions are not always guaranteed in practice. After applying LDA, Eve may seek to verify whether the modification was successful and to estimate the actual attenuation alteration in order to select optimal parameters for the subsequent USD attack. This can be accomplished, in principle, using optical reflectometry techniques \cite{OTDR}, or by directly measuring the intensity of Alice's radiation with a high-sensitivity power meter or a single-photon detector. Once this measurement is performed, Eve can acquire the necessary information to initiate the USD attack and proceed with the USD attack as outlined above.

We have found that
such USD attacks give the whole key to Eve whenever the attenuation
alteration threshold is overcome. It is the 3-photon nature of these
USD attacks that accounts for the existence of the threshold, which
is found to lie in the region of 10--20 dB, depending on the decoy-state
parameters. Such attenuation alterations have been previously experimentally
demonstrated for several types of attenuators under high-power radiation
\cite{LDA_bug2,LDA_mak1}. 

Therefore, one must make sure that the attenuators used in a QKD system are resilient to LDA. One mitigation approach is to monitor the functioning of optical components during the QKD session by placing a beam splitter with a high-sensitivity photodetector at the output of Alice's setup. Another option is to detect external light injection using a watchdog detector, which, however, can be circumvented by Eve \cite{watchdog}. A more promising method is the use of passive monitoring devices, such as optical fuses, which are permanently damaged under high-power radiation, thereby interrupting the communication \cite{lucamarini2015practical}. This effect can be achieved through the fiber-fuse phenomenon \cite{fuse} or by employing carbon nanomaterials \cite{fuse_carbon}. Finally, appropriate types of attenuators should be selected -- those experimentally shown to be more resistant to high-power radiation, such as neutral filters \cite{LDA_bug1} or manual variable attenuators \cite{LDA_mak1}. The actual sustainability of the attenuators used in a particular QKD system should be tested as part of the certification process \cite{LDA_bug2, LDA_mak1, qrate}. In that case, the security of the QKD will be preserved.

\section*{Methods}
\subsection*{Decoy-state method}

The decoy-state QKD utilizes auxiliary coherent states and monitors
their detection statistics. Two decoy states with mean photon numbers
$\nu_{1}$ and $\nu_{2}$ are known to be sufficient for the estimation
of the single-photon fraction. The secret key rate for the
decoy-state BB84 is expressed as follows \cite{key-decoy, ma2005practical}:

\begin{equation}
S = q\,Q_{1}\Big[1-h(E_{1})\Big]-q\,Q_{\mu}\,f(E_{\mu})\,h(E_{\mu}),\label{S}
\end{equation}
where $q$ is the fraction of sifted bits (1/2 for the standard version of BB84), $Q_{1}$ is the single-photon gain, $Q_{\mu}$ is the gain for
signal pulses with mean photon number $\mu$, $E_{1}$ is the quantum
bit error rate (QBER) for the single-photon fraction, $E_{\mu}$ is the overall QBER for the signal pulses, $f(x)$ is the error correction efficiency, $h(x)\equiv-x\log_{2}x-(1-x)\log_{2}(1-x)$
is the binary entropy function. For simplicity, we assume an infinite
key length and the lack of side channels (the security analysis for the decoy-state method in the presence of side channels and finite-key effects can be found in Refs. \cite{navarrete2022improved, ding2023improved}).

Next we consider the optimal vacuum+weak decoy-state variant ($\nu_{1}\equiv\nu$,
$\nu_{2}\equiv0$). The decoy-state method considers an estimation
of $Q_{1}$ and $E_{1}$ given by \cite{ma2005practical}:

\begin{align}
Q_{1} & =  Y_{1}\mu e^{-\mu}\label{Q1}\\
Y_{1} & \geq  Y_{1}^{L}\equiv\frac{\mu}{\nu(\mu-\nu)}\left[Q_{\nu}e^{\nu}-\frac{\nu^{2}}{\mu^{2}}Q_{\mu}e^{\mu}-\frac{\mu^{2}-\nu^{2}}{\mu^{2}}Q_{0}\right]\label{Y1}\\
E_{1} & \leq  E_{1}^{U}\equiv\frac{E_{\nu}Q_{\nu}e^{\nu}-E_{0}Q_{0}}{\nu Y_{1}^{L}}\label{E1},
\end{align}
where $Q_{\mu,\nu,0}$ are the measured gains for signal, decoy and
vacuum states, $E_{\nu,0}$ is the QBER for decoy and vacuum states
correspondingly, $Y_{1}$ is the yield for the single-photon component.

For the gains $Q_{\mu,\nu}$ we have:

\begin{equation}
Q_{\mu,\nu}=\sum_{n=0}^{\infty}p_{\mu,\nu}(n)Y_{n},\label{Q}
\end{equation}
where $p_{\mu,\nu}(n)$ is a Poisson probability function for a photon
number $n$ and the mean value $\mu$ or $\nu$. Here, we have implied
a proper functioning of Alice's apparatus which guarantees that the
desired mean-photon numbers are prepared. The yields for $n$--photon
components $Y_{n}$ can be modified by Eve during PNS, BS or USD attack.
That, however, will be detected by the decoy-state method and reflected
in the single-photon fraction value estimation.

\subsection*{Analytical estimation for USD thresholds}
When the USD attack is performed, the gains are determined by the probability of Eve's conclusive measurement outcome, which is given by the product of three probabilities corresponding to detector clicks at the USD station: two detectors receiving pulses with mean photon number $\tilde{\mu}/4$ or $\tilde{\nu}/4$, and one detector receiving a pulse with mean photon number $\tilde{\mu}/2$ or $\tilde{\nu}/2$. Therefore, the gains for the signal and decoy states are expressed as:

\begin{eqnarray} 
Q_{\mu} & = & \left(1-e^{-\frac{\tilde{\mu}}{2}}\right)\left(1-e^{-\frac{\tilde{\mu}}{4}}\right)^{2}\\
Q_{\nu} & = & \left(1-e^{-\frac{\tilde{\nu}}{2}}\right)\left(1-e^{-\frac{\tilde{\nu}}{4}}\right)^{2}
\end{eqnarray}

Here, we considered perfect noiseless detection at Eve's station and
ideal bit imposition. Such strict requirements, in principle, can be relaxed (See "Pseudo-PNR USD attack simulation results").

Let us derive the critical value for $\kappa$ where the secret key
length estimation starts giving a non-zero value:

\begin{equation}
Y_{1}^{L}(\kappa_{USD})=0\label{eq:eq}
\end{equation}

Considering $Q_0<<1$, eq. (\ref{eq:eq}) takes the following
form:

\begin{equation}
\frac{\Big[1-\exp\Big(-\frac{\tilde{\nu}}{2}\Big)\Big]\Big[(1-\exp\Big(-\frac{\tilde{\nu}}{4}\Big)\Big]^{2}}{\Big[1-\exp\Big(-\frac{\tilde{\mu}}{2}\Big)\Big]\Big[1-\exp\Big(-\frac{\tilde{\mu}}{4}\Big)\Big]^{2}}\frac{\mu^{2}}{\nu^{2}}e^{-(\mu-\nu)}=1
\end{equation}

Note that the assumption on low $Q_0$ value holds even for high channel losses scenario, since the the gains $Q_\mu$ and $Q_\nu$ are fully determined by Eve.

Modifying the left part of the expression we obtain the transcendental
equation for $\kappa_{USD}$:

\begin{equation}
\frac{\sinh(\frac{\kappa_{_{USD}}\nu}{4})\sinh^{2}(\frac{\kappa_{_{USD}}\nu}{8})}{\sinh(\frac{\kappa_{_{USD}}\mu}{4})\sinh^{2}(\frac{\kappa_{_{USD}}\mu}{8})}\frac{\mu^{2}}{\nu^{2}}e^{\frac{\kappa_{_{USD}}-2}{2}(\mu-\nu)}=1\label{eq:tr}
\end{equation}

This equation can be solved numerically. However, for approximate
estimation, we will use the Taylor expansion of the left part:

\begin{equation}
\frac{\sinh(\frac{\kappa_{_{USD}}\nu}{4})\sinh^{2}(\frac{\kappa_{_{USD}}\nu}{8})}{\sinh(\frac{\kappa_{_{USD}}\mu}{4})\sinh^{2}(\frac{\kappa_{_{USD}}\mu}{8})}\lesssim\Big(\frac{\nu}{\mu}\Big)^{3}\label{eq:appr}
\end{equation}

Substituting the right part of the approximation (\ref{eq:appr}) into the transcendental equation (\ref{eq:tr}) we obtain the following equation for $\kappa_{USD}$:

\begin{equation}
\frac{\nu}{\mu}e^{\frac{\kappa_{_{USD}}-2}{2}(\mu-\nu)}=1
\end{equation}

This equation can be solved analytically resulting in equation \eqref{eq:an}.

For the modified USD attack with an additional beam-splitter the gains are modified as follows:

\begin{eqnarray}
Q_{\mu} & = & e^{-(1-T)\tilde{\mu}}\left(1-e^{-\frac{T\tilde{\mu}}{2}}\right)\left(1-e^{-\frac{T\tilde{\mu}}{4}}\right)^{2}\\
Q_{\text{\ensuremath{\nu}}} & = & e^{-(1-T)\tilde{\nu}}\left(1-e^{-\frac{T\tilde{\nu}}{2}}\right)\left(1-e^{-\frac{T\tilde{\nu}}{4}}\right)^{2}
\end{eqnarray}

For the transcendental equation we have:

\begin{equation}
\frac{\sinh(\frac{T\kappa_{_{T}}\nu}{4})\sinh^{2}(\frac{T\kappa_{_{T}}\nu}{4})}{\sinh(\frac{T\kappa_{_{T}}\mu}{4})\sinh^{2}(\frac{T\kappa_{_{T}}\mu}{4})}\frac{\mu^{2}}{\nu^{2}}e^{(\kappa_{T}-1-\frac{T\kappa_{_{T}}}{2})(\mu-\nu)}=1
\end{equation}

Using approximation (\ref{eq:appr}) we found an estimate for $\kappa_{T}$ resulting in equation \eqref{kappa_T}.

\subsection*{3-photon PNS attack}
Let us consider a specific PNS strategy with blocking all $n$--components
with $n\neq3$. This strategy leads to a more significant drop in
secret key rate in comparison to the standard PNS strategy with preserving
$n=2$ component. This 3-photon attack, however, does not require
a quantum memory as the measurements in two different bases can be
conducted simultaneously. Hence, only classical memory for the measurements
outcome is required. 

We consider the zero yields for 3-photon attack, except for the $Y_{3}$:

\begin{eqnarray}
Y_{n\neq3} & = & 0\\
Y_{n=3} & \neq & 0
\end{eqnarray}

LDA modifies mean photon numbers as dictated by eq.~(\ref{eq:mu_LDA}) and (\ref{eq:nu_LDA}). The gains are then:

\begin{eqnarray}
Q_{\mu} & \approx & Y_{3}\frac{\tilde{\mu}^{3}}{6}e^{-\tilde{\mu}}\\
Q_{\nu} & \approx & Y_{3}\frac{\widetilde{\nu}^{3}}{6}e^{-\tilde{\nu}}
\end{eqnarray}

The critical value for $\kappa$ where the secret key
length estimation starts giving non-zero value can be derived from the following equation:

\begin{equation}
Y_{1}^{L}(\kappa_{3ph})=0
\end{equation}

Considering $p_{dcr}<<1$, we obtain the equation:

\begin{equation}
Q_{\nu}e^{\nu}=\frac{\nu^{2}}{\mu^{2}}Q_{\mu}e^{\mu},\label{eq:3ph}
\end{equation}

Solving equation (\ref{eq:3ph}) for $\kappa_{3ph}$ we obtain the
3-photon attack threshold:

\begin{equation}
\kappa_{3ph}=1+\frac{1}{\mu-\nu}\ln\frac{\mu}{\nu}
\end{equation}

The threshold is proportional to those for the described USD attacks \eqref{eq:an} and \eqref{kappa_T}. This underlines the 3-photon nature of the described USD strategies.

\subsection*{Simulation of USD attack}
Considering noiseless and lossless detection at Eve's station, we
can express the yields for the USD attack, i.e. the probabilities
of conclusive measurements outcome for $n$--photon components. For
this, we use the binomial distribution of photons after the 50:50
beam-splitter and pick up the scenarios, when only 3 detectors click
(we assume Eve using threshold detectors without photon number resolution):

\begin{align}
Y_{n}^{USD} & =  \sum_{k=1}^{n}\binom{n}{k}\Big(\frac{1}{2}\Big)^{k}\Big(\frac{1}{2}\Big)^{n-k}\nonumber &\\
 & \times  \sum_{m=1}^{n-k-1}\binom{n-k}{m}\Big(\frac{1}{2}\Big)^{k}\Big(\frac{1}{2}\Big)^{n-k-m}=1-\frac{3^{n}-1}{2^{2n-1}},\label{Y}\\
\nonumber 
\end{align}
where the first sum denotes the probability of $k$ photons to arrive
at the correct-basis measurement side and the second sum denotes the
probability of the the remaining $n-k$ photons to be divided along
two paths in the incorrect-basis measurement side.

For the modified USD attack the yields are multiplied by the probability
of $n$ photons passing through the beamsplitter $T$:

\begin{equation}
Y_{n}^{T}=T^{n}\cdot Y_{n}^{USD}\label{Y_T}
\end{equation}

We have simulated the single-photon gain estimation and the total gain for signal pulses using equations \eqref{Q1}--\eqref{Y1}, \eqref{Q}, \eqref{Y}--\eqref{Y_T} for different decoy-state parameters. The series in Eq.~\eqref{Q} was truncated at $n = 600$, since the cumulative probability of the Poisson distribution beyond this point becomes negligible within the precision of the 8-digit floating-point arithmetic used in our simulations. The results are presented in "USD attacks simulation results."

\subsection*{Simulation of Pseudo-PNR USD attack}

The purpose of PPNR USD attack is to match all four parameters monitored by Bob, which in absence of Eve are given by:
\begin{eqnarray}
Q_\mu &\approx& \mu\eta\eta_B + p_{dcr},\label{Q_mu}\\
Q_\nu &\approx& \nu\eta\eta_B + p_{dcr},\label{Q_nu}\\
Q_\mu^{(2)} &\approx& (\frac{\mu\eta\eta_B}{2})^2 + \mu\eta\eta_B\,p_{dcr} + p_{dcr}^2,\label{Q2_mu}\\
Q_\nu^{(2)} &\approx& (\frac{\nu\eta\eta_B}{2})^2 + \nu\eta\eta_B\,p_{dcr} + p_{dcr}^2.\label{Q2_nu}
\end{eqnarray}

By performing PPNR USD attack, Eve modifies the gains as follows:

\begin{equation}
\begin{alignedat}{1}
    Q_{\mu} = Q_{\mu}^{USD} \Big(P_\mu(0)\,p_{\nu,1} &+ P_\mu(1)\,p_{\nu'} +\\ P_\mu(2)\,p_{\mu,1} &+ P_\mu(3)\,p_{\mu'}\Big) + p_{dcr},\label{Q_mu_E}
\end{alignedat}
\end{equation}\begin{equation}
\begin{alignedat}{1}
    Q_{\nu} = Q_{\nu}^{USD} \Big(P_\nu(0)\,p_{\nu,1} &+ P_\nu(1)\,p_{\nu'} +\\ P_\nu(2)\,p_{\mu,1} &+ P_\nu(3)\,p_{\mu'} \Big)+ p_{dcr},\label{Q_nu_E}
\end{alignedat}
\end{equation}\begin{equation}
\begin{alignedat}{1}
    Q_{\mu}^{(2)} = Q_{\mu}^{USD} \Big(P_\mu(0)\,p_{\nu,1}\,p_{dcr} &+ P_\mu(1)\,p_{\nu'}^{(2)} +\\ P_\mu(2)\,p_{\mu,1}\,p_{dcr} &+ P_\mu(3)\,p_{\mu'}^{(2)} \Big)+p_{dcr}^2,\label{Q2_mu_E}
\end{alignedat}
\end{equation}\begin{equation}
\begin{alignedat}{1}
    Q_{\nu}^{(2)} = Q_{\nu}^{USD} \Big(P_\nu(0)\,p_{\nu,1}\,p_{dcr} &+ P_\nu(1)\,p_{\nu'}^{(2)} + \\P_\nu(2)\,p_{\mu,1}\,p_{dcr} &+ P_\nu(3)\,p_{\mu'}^{(2)} \Big)+p_{dcr}^2,\label{Q2_nu_E}
\end{alignedat}
\end{equation}
where 
\begin{eqnarray}
Q_{\mu}^{USD}& = & \left(1-e^{-\frac{T\tilde{\mu}\eta_E}{2}}\right)\left(1-e^{-\frac{T\tilde{\mu}\eta_E}{4}}\right)^{2},\\
Q_{\text{\ensuremath{\nu}}}^{USD} & = & \left(1-e^{-\frac{T\tilde{\nu}\eta_E}{2}}\right)\left(1-e^{-\frac{T\tilde{\nu}\eta_E}{4}}\right)^{2}
\end{eqnarray}
are the probabilities of a conclusive USD outcome for the signal and the decoy state, respectively;
\begin{flushleft}
\begin{align}
P_\mu(k) &= \binom{3}{k} \!\left(1 - e^{-\frac{(1 - T)\tilde{\mu}\eta_E}{3}}\right)^{\!k}
           \!\left(e^{-\frac{(1 - T)\tilde{\mu}\eta_E}{3}}\right)^{\!1-k},\\
P_\nu(k) &= \binom{3}{k} \!\left(1 - e^{-\frac{(1 - T)\tilde{\nu}\eta_E}{3}}\right)^{\!k}
           \!\left(e^{-\frac{(1 - T)\tilde{\nu}\eta_E}{3}}\right)^{\!1-k}
\end{align}
\end{flushleft}
are the binomial probabilities of a $k$-photon count on the PPNR detector for the signal and the decoy state, respectively;
\begin{eqnarray}
p_{\mu'} &=& 1 - e^{-\mu' \eta_B},\label{p_mu}\\
p_{\nu'} &=& 1 - e^{-\nu' \eta_B}\label{p_nu}
\end{eqnarray}
are the probabilities of a single click at Bob's station induced by a laser pulse sent by Eve with a mean photon number $\mu'$ and $\nu'$, respectively;
\begin{eqnarray}
p_{\mu'}^{(2)} &=& \left(1 - e^{-\frac{\mu' \eta_B}{2}}\right)^2,\label{p2_mu}\\
p_{\nu'}^{(2)} &=& \left(1 - e^{-\frac{\nu' \eta_B}{2}}\right)^2\label{p2_nu}
\end{eqnarray}
are the probabilities of a double click at Bob's station induced by a laser pulse sent by Eve with a mean photon number $\mu'$ and $\nu'$, respectively; $p_{\mu, 1}$, $p_{\nu, 1}$ are the probabilities of a single click at Bob's station induced by a single photon sent by Eve after 2- and 0-photon click at PPNR detector, respectively.

By equating expressions (\ref{Q_mu})--(\ref{Q2_nu}) to (\ref{Q_mu_E})--(\ref{Q2_nu_E}) and combibing the resulting equations, we obtain two systems of linear equations. The first one is:
\begin{equation}
\begin{alignedat}{1}
    &Q_{\mu}^{USD}\,P_\mu(1)\Big(p_{\nu'}^{(2)} - p_{\nu'}\,p_{dcr}\Big) + \\
    &Q_{\mu}^{USD}\,P_\mu(3)\Big(p_{\mu'}^{(2)} - p_{\mu'}\,p_{dcr}\Big) = (\frac{\mu\eta\eta_B}{2})^2,\label{Q2_mu_eq}
\end{alignedat}
\end{equation}\begin{equation}
\begin{alignedat}{1}
    &Q_{\nu}^{USD}\,P_\nu(1)\Big(p_{\nu'}^{(2)} - p_{\nu'}\,p_{dcr}\Big) + \\
    &Q_{\nu}^{USD}\,P_\nu(3)\Big(p_{\mu'}^{(2)} - p_{\mu'}\,p_{dcr}\Big) = (\frac{\nu\eta\eta_B}{2})^2.\label{Q2_nu_eq}
\end{alignedat}
\end{equation}

By solving this system using Cramer's rule, we express the linear relation between $p_{\mu'}$ and $p_{\mu'}^{(2)}$ and between $p_{\nu'}$ and $p_{\nu'}^{(2)}$. Using explicit expressions (\ref{p_mu})--(\ref{p2_nu}) we express $\mu'$ and $\nu'$.

The second system of linear equations is

\begin{equation}
\begin{alignedat}{1}
    Q_{\mu}^{USD} \Big(P_\mu(0)\,p_{\nu,1} + P_\mu(2)\,p_{\mu,1}\Big) = &\\ \mu\eta\eta_B - Q_{\mu}^{USD}\,P_\mu(1)\,p_{\nu'} -\, & Q_{\mu}^{USD}\,P_\mu(3)\,p_{\mu'},\label{Q_mu_eq}
\end{alignedat}
\end{equation}\begin{equation}
\begin{alignedat}{1}
    Q_{\nu}^{USD} \Big(P_\nu(0)\,p_{\nu,1} + P_\nu(2)\,p_{\mu,1}\Big) = &\\ \nu\eta\eta_B - Q_{\mu}^{USD}\,P_\nu(1)\,p_{\nu'} -\, & Q_{\mu}^{USD}\,P_\nu(3)\,p_{\mu'}.\label{Q_nu_eq}
\end{alignedat}
\end{equation}

This system can be also solved for $p_{\mu,1}$ and $p_{\nu,1}$ using Cramer's rule and substituting the previously found $\mu'$ and $\nu'$.

For the solution to be valid, the following conditions must be satisfied: 
\begin{eqnarray}
p_{\mu, 1} \in [0, \eta_E]&;\,
p_{\mu'} \in [0, 1]&;\,
p_{\mu'}^{(2)} \in [0, 1];\,\\
p_{\nu, 1} \in [0, \eta_E]&;\,
p_{\nu'} \in [0, 1]&;\,
p_{\nu'}^{(2)} \in [0, 1].
\end{eqnarray}
These constraints cannot be satisfied for arbitrary $\kappa$. The solution becomes consistent only above a certain threshold value. Eve, however, does not necessarily need to satisfy the equations exactly. Alice and Bob would likely not notice if the observed statistics deviated from the expected values by a small margin. To determine the threshold in this case, we numerically search for the smallest $\kappa$ at which the ratios of observable to expected gains equal unity within a nonzero tolerance. To this end, we set Eve's parameters $\mu'$, $\nu'$, $p_{\nu,1}$, and $p_{\mu,1}$ according to the closed-form solution, clipping their values within the constraint interval, and perform a sweep over $\kappa$ and the beamsplitter transmittance $T \in [0,1]$ using the golden-section search.

We have peformed numerical simulations to find threshold values for attenuation alteration $\kappa_{PPNR}$ for PPNR USD attack. We have employed different decoy-state parameters, channel losses, Eve's detection efficiency and the dark-count rates of Bob's detectors. The results are presented in "Pseudo-PNR USD attacks simulation results."

\section*{Data availability}
All the data that support the results of the current
work are available from the corresponding authors upon
reasonable request.

\section*{Code availability}
The code for simulation is available from the corresponding author upon reasonable request.

\section*{Acknowledgments}

I.S.S. thanks S.P. Kulik for numerous discussions. I.S.S. also acknowledges the scholarship from the Foundation for the Advancement of Theoretical Physics and Mathematics "BASIS".

\section*{Funding}
No funding was received for conducting this study.

\section*{Author contribution}

I.S. developed the theoretical framework and wrote the manuscript, 
I.S. and K.B. introduced the scenario for combined LDA and USD attack, 
I.S. and R.L. executed the numerical simulations,  
D.B., A.S., V.V. and D.M. aided in interpreting the results and worked on the manuscript,
S.M. and D.D. conceptualized and supervised the work.

\section*{Competing Interests}
The authors declare no competing interests.

\end{document}